# Mechanistic forecasts of species responses to climate change: the promise of biophysical ecology

**Authors:** Natalie Briscoe*, Shane Morris*, Paul Mathewson, Lauren Buckley, Marko Jusup, Ofir Levy, Ilya Maclean, Sylvain Pincebourde, Eric Riddell, Jessica Roberts, Rafael Schouten, Mike Sears & Michael Kearney

*contributed equally

## Abstract

A core challenge in global change biology is to predict how species will respond to future environmental change and to manage these responses. To make such predictions and management actions robust to novel futures, we need to accurately characterize how organisms experience their environments and the biological mechanisms by which they respond. All organisms are thermodynamically connected to their environments through the exchange of heat and water at fine spatial and temporal scales and this exchange can be captured with biophysical models. Although mechanistic models based on biophysical ecology have a long history of development and application, their use in global change biology remains limited despite their enormous promise and increasingly accessible software. We contend that greater understanding and training in the theory and methods of biophysical models is vital to expand their application. Our review shows how biophysical models can be implemented to understand and predict climate change impacts on species' behavior, phenology, survival, distribution, and abundance. It also illustrates the types of outputs that can be generated, and the data inputs required for different implementations. Examples range from simple calculations of body temperature at a particular site and time, to more complex analyses of species' distribution limits based on projected energy and water balances, accounting for behavior and phenology. We outline challenges that currently limit the widespread application of biophysical models relating to data availability, training, and the lack of common software ecosystems. We also discuss progress and future developments that could allow these models to be applied to many species across large spatial extents and timeframes. Finally, we highlight how biophysical models are uniquely suited to solve global change biology problems that involve predicting and interpreting responses to environmental variability and extremes, multiple or shifting constraints, and novel abiotic or biotic environments.

## Introduction

Accurate forecasts of how environmental change will affect species are vital if we are to effectively manage biodiversity now and in the future. Yet predicting how organisms respond to environmental change is complex because such responses are generally non-linear, often have thresholds, and may change with novel conditions (Beissinger & Riddell, 2021; Huey et al., 2012). Thus, there is growing recognition that we need to explicitly incorporate mechanisms into models of species' responses to environmental change if we are to improve predictions and better manage outcomes (Helmuth et al., 2005; Keith et al., 2008; Urban et al., 2016).

Exactly what mechanisms to incorporate is a daunting question as they could relate to most topics in ecology, evolution, and physiology, such as life history, population dynamics, dispersal, and biotic interactions (Briscoe et al., 2019; Ehrlén & Morris, 2015; Thuiller et al., 2013). A useful starting point is to model fundamental constraints on fitness such as survival, development, growth, and reproduction. Models based on the principles of biophysical ecology (hereafter *biophysical models*) capture the balances of heat, water, or mass exchange between organisms and their microclimatic environment and translate these into metrics of performance (Figure 1), offering a conceptually simple way to capture the fundamental physical and chemical constraints relevant to all living things (Gates, 1980). Thus, they are a judicious starting point in analyses of how environmental changes – particularly in climate – will affect organisms. Biophysical models also often form the basis of “mechanistic niche models” (also referred to as “ecophysiological” or “mechanistic” models), which can incorporate additional processes (e.g., metabolic theory, demographic, evolutionary).

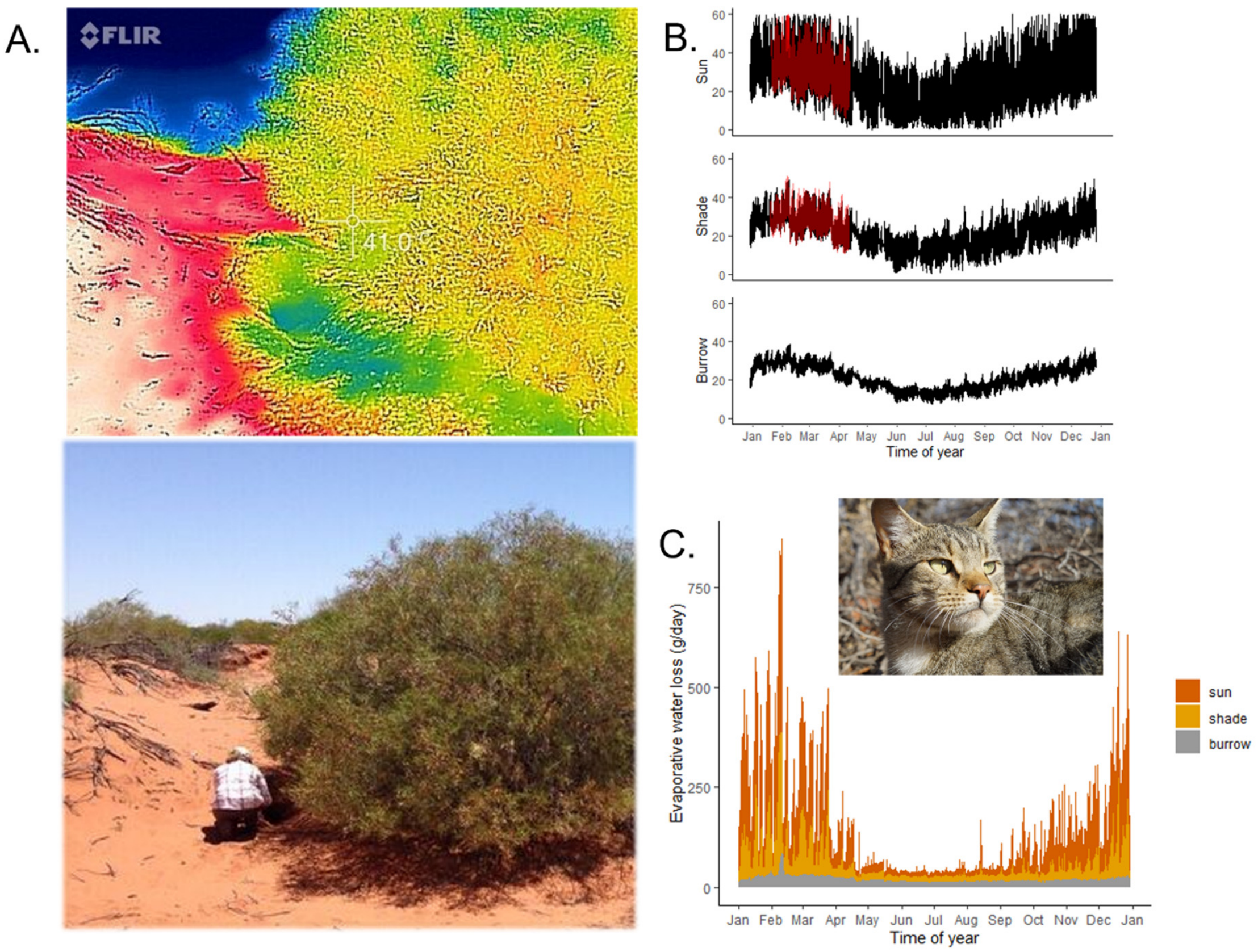

**Figure 1. Biophysical models are powerful tools for capturing how an organism’s environment affects its physiological condition. A. Thermal image showing the variation in surface temperatures on a sand dune at a site in arid Australia, B. Hourly temperatures in microclimates available to feral cats (surface temperatures in the sun, shade and down a burrow) at the site, as modelled by a microclimate model using the principles of biophysics (black) and measured using temperature loggers (red) C. Predicted daily water costs of feral cats (image: Hugh McGregor) using each**

**microclimate (red = surface (sun), orange = surface (shade), grey = burrow). Costs were estimated using a biophysical model parameterized using data on feral cat functional traits (Briscoe et al., 2022).**

The principles of biophysical ecology have a long history of application to the study of adaptations of organisms (Porter & Gates, 1969) and are also incorporated into larger-scale models of climate hydrology and vegetation models (Maclean et al., 2015; Michaletz et al., 2016). Despite their enormous promise, biophysical models are not yet routine practice in studies seeking to predict species responses to global change. Thirty years ago, O'Connor and Spotila (1992) recognised the slow uptake of biophysical methods in ecology. These models have since become more sophisticated and accessible, and the need for their predictions and inferences has only become greater.

Here we first review and outline biophysical models, focusing on how they differ from statistical models, the different ways they can be implemented, and the types of questions they can be used to answer. We draw on our own experience, as well as a literature review of how these models have been applied to animals (Supporting Information). We focus predominantly on terrestrial animal studies but the methods are relevant to plants (including in ecosystem models) aquatic organisms, and humans (Campbell-Staton et al., 2021; Muir, 2019; Sarà et al., 2011). Second, we highlight limitations that hamper broader use of biophysical models such as training, data and disciplinary divisions, and discuss the progress that has been made and future opportunities.

## Mechanistic versus statistical models

Most models used in ecology are statistical or 'phenomenological' in nature (Figure 2a), directly describing the observed patterns or relationships between predictors and phenomena of interest (Connolly et al., 2017). In contrast, mechanistic models predict a phenomenon of interest based explicitly on one or more underlying processes. While acknowledging that mechanistic and statistical approaches described here represent either end of a continuum (Dormann et al., 2012), an appreciation of their differences is a useful starting point for understanding biophysical models and their potential contribution to global change biology.

Statistical approaches start with the data. When fitting these models, the strategy is to find relationships between the phenomena of interest and predictor variables, but with underlying processes left implicit so that the data lead the dance (Hilborn & Mangel, 1997). Thus, major challenges lie in the choice of models and predictors and there is a strong emphasis on uncertainty and error propagation, and model-data fusion and feedback (Dietze, 2017). The flexibility of statistical approaches means that they can be applied to a broad range of problems without explicit knowledge of the likely constraints on the system (Dormann et al., 2012).

In contrast, mechanistic approaches start by assuming that a particular set of processes are influencing the phenomena of interest. In the case of biophysical models, the strategy is to start with fundamental processes relating to energy and mass exchange between an individual (the system) and its surroundings (the environment) and use the outcomes as the basis for inferring survival, growth, development and reproduction (Kearney & Porter, 2009). These outputs can be integrated with other types of models, for example those focused on capturing demography or movement (Buckley et al., 2010; Sears et al., 2016). In mechanistic approaches, the underlying theory of the modelled processes leads the dance and tightly constrains the choice of models and their associated parameters and predictor variables. The major challenges lie in balancing realism vs. abstraction of the models to be used, and in obtaining the parameters and predictor variables. This balancing of realism and abstraction in biophysical models requires a deep understanding of both the natural history and the underlying physical theory, which is an increasingly rare outcome of biological training in ecology (Bialek & Botstein, 2004; Hampton & Wheeler, 2012).

**BOX: Formal distinctions between statistical and mechanistic models**

A statistical model assumes that a dataset generated by the phenomenon of interest contains realisations of a random variable drawn from a particular distribution. This distribution is characterised by its parameters, such as a mean $\mu$ and a variance $\sigma$. Subsequently, distribution parameters are modelled as functions of one or more environmental predictor variables **x**, for example,

$$\mu_i = g^{-1}(\beta_1 x_i^1 + \ldots + \beta_n x_i^n),$$

in the case of generalised linear models, where $g$ is the 'link function'. The model parameters $\boldsymbol{\beta}$ (boldface means a vector) —the effect of each variable $x$ on the phenomenon of interest $\mu$— can be estimated by finding those that maximise the likelihood of observing the dataset **y**, where the likelihood function is given by the initially assumed distribution.

In contrast, mechanistic models take the form

$$\frac{\mathrm{d}\mathbf{x}}{\mathrm{d}t} = \mathbf{f}(\mathbf{x}, t, \mathbf{u}; \boldsymbol{\beta}),$$

where a vector of system state variables **x** (e.g., an individual's body temperature and water balance) is predicted through time $t$ as a function of a vector of exogenous forcings $\boldsymbol{u}$ (e.g., radiation, wind speed, humidity, air temperature), where $\boldsymbol{\beta}$ is a vector of model parameters (e.g., surface area, body insulation, solar absorptivity). The vector function **f** is a collection of physical laws in functional form, one for each state variable in **x** (e.g., the processes of convective, radiative and evaporative heat transfer). The parameters **β** are, in general, estimable via the maximum likelihood method applied to the measurements y, after establishing a functional relationship between y and the state variables x, y=F(x), and assuming a probability distribution for measurement errors.

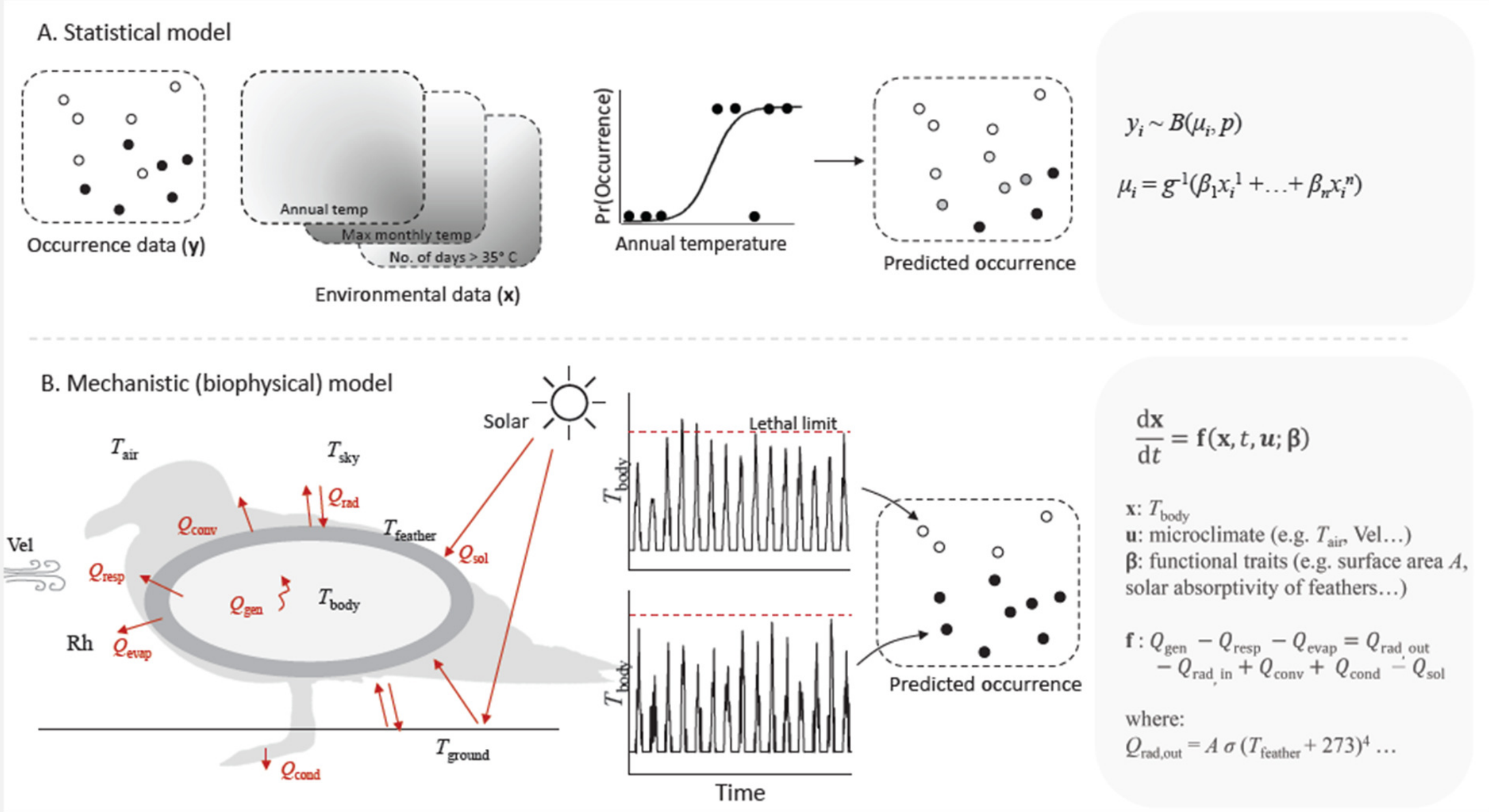

**Figure 2. Overview of components and decisions made when building (A) statistical vs. (B) mechanistic (biophysical) models to predict a phenomenon of interest (here: organismal occurrence, μ). Statistical models use relevant environmental covariates (*x*: here relating to temperature) using observations of that phenomenon (*y*). When fitting the model, key decisions include the assumed probability distribution of the data (here a binomial distribution), covariates to include, and the shape of the modelled response. In contrast, mechanistic models describe the phenomenon of interest by simulating underlying processes (here: overheating and reaching a lethal body temperature). First, the model calculates the body temperature of the organism given its surrounding**

**microclimate (via radiation, convection, conduction, metabolic heat generation and evaporation) and its traits (e.g., solar absorptivity of feathers, surface areas, basal metabolic rate, behavioral and physiological regulation options and parameters). Next, the model can predict the risk of overheating by comparing the calculated body temperatures to the lethal body temperature of the organism. This can then be used to infer occurrence. Key decisions typically relate to simplifying assumptions. Here, the bird is assumed to approximate an ellipsoid, be in the sun (full solar radiation) and on the ground.**

## Advantages of biophysical models for predicting, attributing, and understanding impacts of climate change

To date, studies of how global change will affect species have predominantly employed statistical approaches, but there is a growing demand for mechanistic approaches that can generate more reliable predictions under novel future conditions and identify key drivers of change and management levers (Buckley et al., 2010; Urban et al., 2016). Our literature review (see Supporting Information) identified 211 papers that have applied biophysical models to animals, the majority of these (65%) since 2010 (Figure 3a). Biophysical modelling applications were initially biased to ectotherms, but are now also used for endotherms, with their application to both groups increasing. Overall, 36% of studies (a total of 76 papers) we identified modelled species responses to past or future climate change or discussed model applications in the context of climate change; this rises to 50% when only studies from 2010 are considered. Despite the limited applications of biophysical models to climate change studies so far, several important insights are already emerging and give a sense of what we could learn if their application was broadened.

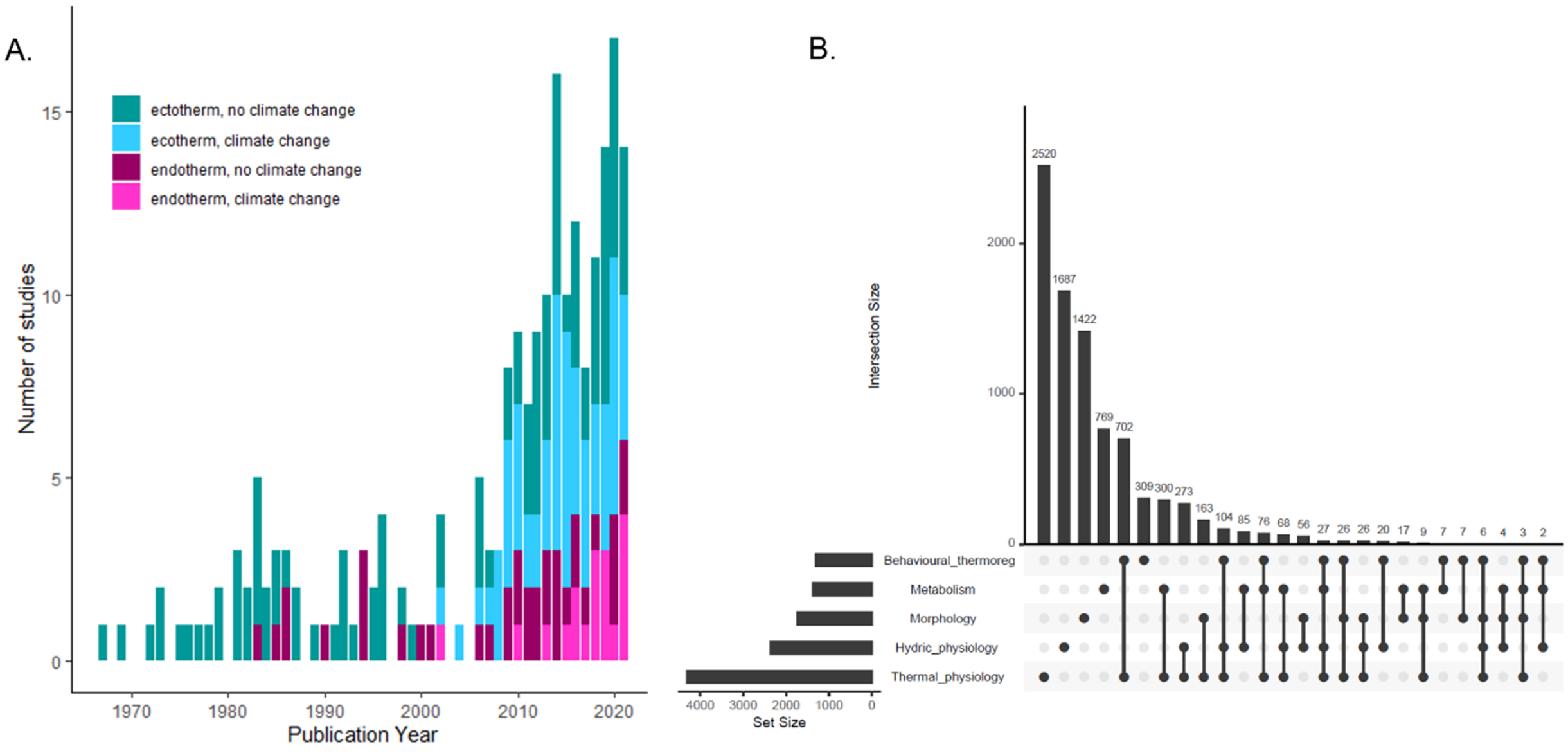

**Figure 3. (a) Number of studies per year that applied biophysical models to animals, showing type of taxa (ectotherms/endotherms), and whether the study considered climate change responses, (b) Number of studies focused on lizard and snakes identified using keywords related to relevant functional traits related to thermal physiology, hydric physiology, morphology, metabolism or behavioral thermoregulation. The bottom (left) histogram indicates the total number of studies identified in each search, while the top panels display the number of studies in each set of terms, as indicated by the filled circles below the x-axis.**

The great power of biophysical models is that they can be used to infer what will happen under any combination of functional traits and environment because they are based on universal physical principles. Thus, they could, in theory, predict the body temperature of an organism on another planet if we knew the environment there. As a result, biophysical models can make confident predictions of the consequences of novel climates for species given their functional biophysical traits. As organisms will increasingly be exposed to novel conditions under climate change (e.g., more extreme conditions, and new combinations of climate) (Davy et al., 2017), the predictive ability of statistical models may erode because no observations under these conditions (Box 1) are available to parameterise such models (Buckley et al., 2010; Sinclair et al., 2010). In contrast, biophysical models inherently translate environmental conditions through time into currencies directly relevant to the fitness of the organism and allow new processes not yet captured in observations to become limiting as conditions change.

The fact that biophysical models can capture limiting factors makes them ideal tools for attributing observed shifts in distribution, phenology, population dynamics or behavior of a species to climate change (Kearney, et al., 2010a; Riddell et al., 2019). They can also reveal management levers (e.g., shade manipulation, water or food provisioning, translocations) for adapting to climate change impacts (Mitchell et al., 2008, 2013).

## Biophysical models – a brief overview

At the core of biophysical models are equations for the exchange of energy and mass between an organism and its environment (Figure 4). These models consider the organism as a thermodynamic system, where incoming energy must equal outgoing energy plus any energy stored (see books by Gates (1980) and Campbell and Norman (1998) or O'Connor and Spotila (1992) for a shorter overview). A useful analogy is the balancing of a bank account, where one must account for various streams of income and types of expenses. For example, a bird (Figure 2) or lizard (Figure 5) on the open ground will gain energy from the environment as heat from direct, scattered and reflected solar radiation, as well as infrared radiation from the sky, ground, and vegetation. They will produce metabolic heat and they will also lose heat through infrared radiation and evaporation of water from their surface and via respiration. Heat exchange via contact with solid surfaces such as the ground (i.e., conduction) or immersion in air or fluid (i.e., convection) can be gains or losses depending on the temperature gradient between the organism and its surroundings. All these factors eventually determine the thermal energy of the organism's body, manifested as its body temperature.

In biophysical models, these heat exchange process account for both the environment and the traits of the organisms (Barlett & Gates, 1967). For example, the solar radiation absorbed by the lizard depends on the incoming solar radiation (perhaps mediated by shade from plants or terrain), the surface areas exposed and the absorptivity of these surfaces. Convective heat exchange depends on the temperature difference between the lizard's surface and the air, the surface area exposed to the air, the lizard's size and shape, and the properties of the air (e.g., temperature, density, velocity). The same principles apply to heat exchange for any other type of organism though the dominating processes and required functional traits may vary. The heat balance equation in Figure 4 can be solved for steady-state body temperature or for metabolic heat production and provides estimates of evaporative heat loss of an organism in a particular environment. These outputs are a powerful starting point for making inferences about how the environment constrains the species' distribution, behavior and phenology (Figure 5; Table 1).

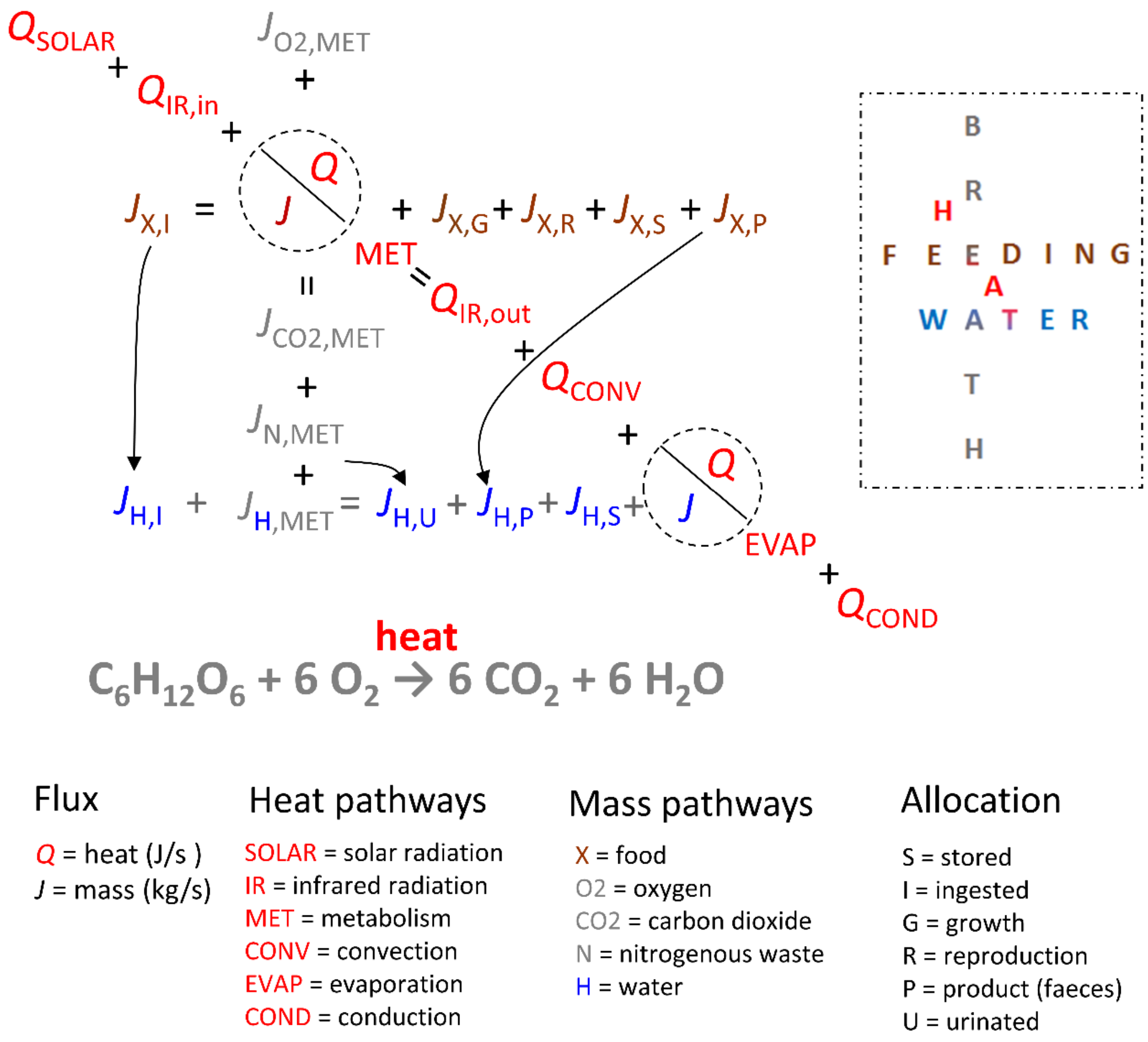

**Figure 4. The coupled equations for the exchange of energy and mass between an organism and its environment via heat (red), respiration (grey), feeding (brown) and water (blue). At the core of biophysical models is the heat budget (diagonal) equation that calculates the energy exchanged through conduction, convection, radiation, evaporation, and metabolism. It is intersected by the mass balance equations for allocation of energy from food (horizontal equation) and respiration (vertical equation) at the metabolism term and water at the respiration term.**

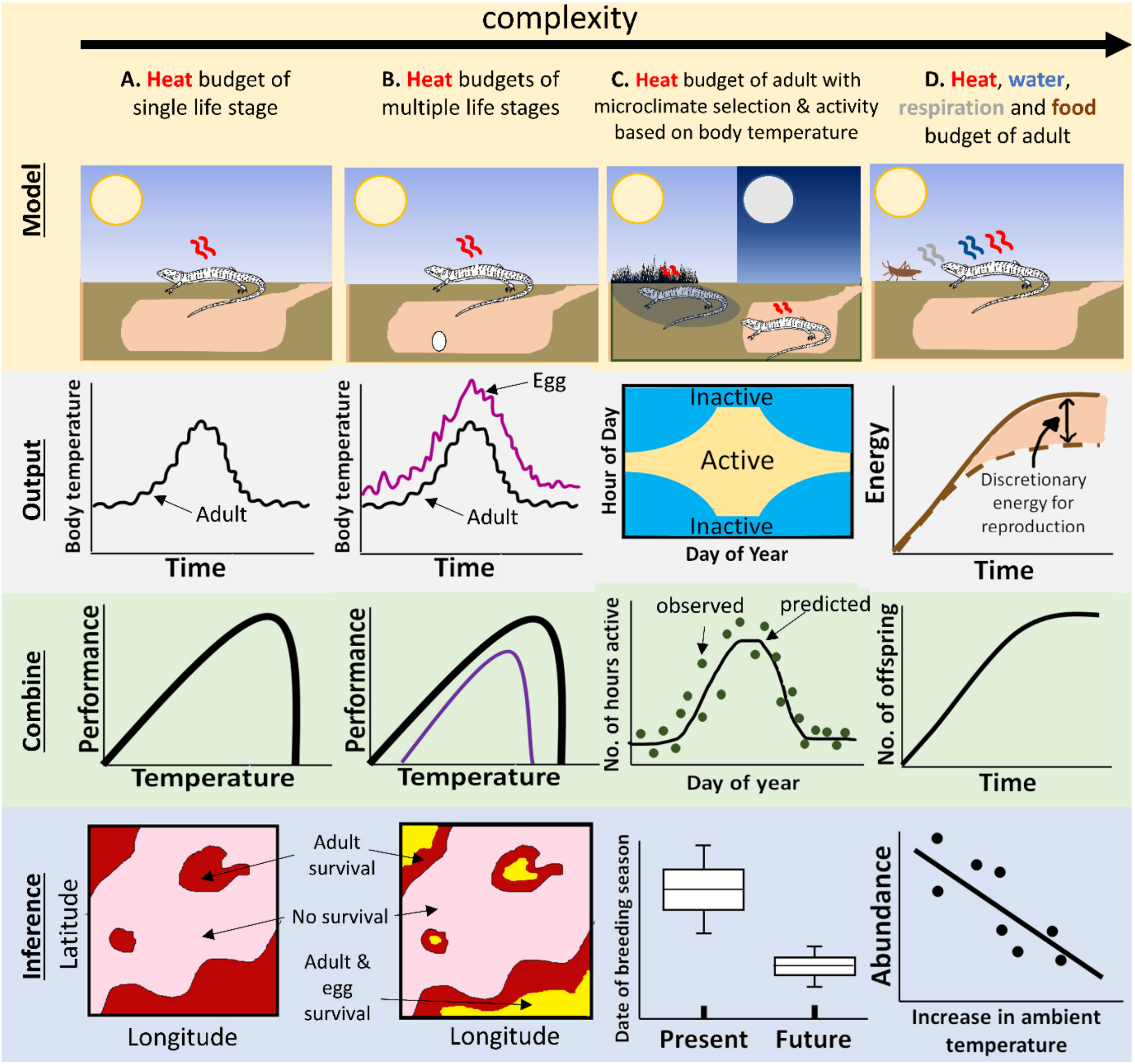

**Figure 5. Biophysical models of a lizard that vary in modelled outputs, mechanisms and complexity (A-D). (A) The heat exchange equation (red, see Figure 4) is used to predict the body temperature of adult (or adult and egg (B)) life stages and combined with thermal performance curves to infer distribution limits. (C) Predictions of the lizard's body temperature are used to predict potential activity times and infer shifts in phenology under climate change. (D) The entire energy and water budget of the lizard is calculated using the energy and mass balance equations (see Figure 4) and foraging activity is constrained by predicted body temperatures and desiccation risk. Calculations of discretionary energy are used to infer potential reproductive output and predict population growth through time.**

In addition to heat exchange, organisms exchange energy through work (e.g., movement) and mass (food). To determine whether the animal can grow and reproduce, we can extend our analysis to consider its entire energy and water budget using coupled energy and mass balance equations that capture the exchange of food, water, and respiratory gases and metabolic waste (Porter & Tracey (1983), Figure 4). Energy available for metabolism, growth, and reproduction can be calculated using information on the energy density and amount of food ingested and the proportion of this lost in faeces or to microflora (Buckley, 2008; Levy et al., 2017). The dynamics of metabolic processes can be calculated with metabolic theory (Kooijman, 2010), a large topic that is beyond the scope of this review (but see Kearney et al., 2010b, 2013, 2021). The characteristics of food ingested also determine water gained from food and lost via faeces, while the water balance is influenced by metabolic water and nitrogenous waste produced via metabolism, as well as that lost via evaporation.

The coupling of the energy and mass balance (Figure 4) reduces the degrees of freedom of the overall problem and highlights how the organism is an interconnected system with inherent feedbacks and trade-offs. For example, a lizard foraging on the ground over summer may be subject to high radiant heat loads, requiring high rates of evaporative water loss if it is to avoid hyperthermia (Loughran & Wolf, 2020). The lizard can avoid these water costs by ceasing activity and sheltering in shade or in a burrow, but this simultaneously reduces food intake (Buckley, 2008; Kearney et al., 2009a; Levy et al., 2017). The vulnerability of the lizard to reduced food intake or enhanced water loss will depend on its recent history of feeding and heat stress, with consequences for its future growth and reproduction, emphasising the importance of the temporal context.

The focus on individuals in biophysical modelling allows a strong connection between theory and observation; model parameters and predictions (e.g., of body temperatures, activity, microhabitat use, energy and water turnover) can be directly observed and measured (Briscoe et al., 2022; Kearney et al., 2018; Mathewson et al., 2020). An expedient strategy can be to start with simple biophysical models that broadly bound the problem and then add complexity as required to adequately account for observations (e.g., Porter et al., 1973). Biophysical models also generate predictions and explanations that can be tested at different scales and levels of organisation. For example, models can be used to predict how active individuals can be at a particular site at a particular time (Levy et al., 2012), how foraging activity determines reproductive output across sites and years (Adolph & Porter, 1993; Kearney, 2012), and how this in turn drives population dynamics and distribution limits (Buckley, 2008). By varying model parameters in sensitivity analyses one can generate hypotheses about the strength of selection on trait values (Kearney et al., 2009b) and predict clines (Sears & Angilletta, 2004).

An important consideration when using biophysical models for global change biology is whether relative outputs (e.g., indices) are sufficient, or whether more accurate estimates of the organism's state are required (O'Connor & Spotila, 1992). While relative metrics are sufficient for some applications (e.g., identifying regions likely to experience the largest increases in cooling costs), accuracy is often necessary when identifying hard limits on where the species can occur – for example sites where individuals would exceed lethal body temperatures or be unable to meet their energy or water requirements.

One of the challenges in biophysical models is to accurately specify the environment experienced by the organism through detailed measurements or microclimate models (Figure 1). Biophysical models demand accurate estimates of specific aspects of the microclimates experienced by organisms that directly influence the heat balance (Pincebourde & Woods, 2020) (e.g., wind speed and solar radiation in addition to humidity and air temperature, Figure 2b) at scales relevant to the organism – usually meters or finer and hours (Potter et al., 2013). This information can be measured directly (e.g., with portable weather stations (Briscoe et al., 2014), thermal cameras (Choi et al., 2019), and temperature loggers (Lembrechts et al., 2020; Maclean et al., 2021) or translated form gridded or weather station data using microclimate models (Figure 1) (Kearney & Porter, 2017; Maclean et al., 2019; Porter et al., 1973).

A related challenge lies in capturing and simulating how animals use the microclimates available to them. Microclimates can vary across an organism's habitat by as much as 20-30°C depending on topographical and vegetational features (Bakken, 1989; Sears et al., 2011). Studies often characterise the microhabitats used by the species (full sun, full shade, burrow; Figure 1) and assume that animals can select between these options to avoid lethal conditions or remain as close as possible to preferred temperatures at any point in space (e.g., Buckley et al., 2010, Kearney et al., 2018). However, depending on the spatial distribution of these temperatures, animals may or may not have access to suitable temperatures (Sears & Angilletta, 2015). The issue of accessibility is especially important to small animals whose body temperatures change rapidly in response to local thermal microenvironments (Sears & Angilletta, 2015; Stevenson, 1985a). Biophysical models that integrate movement and thermoregulation are a promising approach to understand how spatial heterogeneity can impact the activity and energetics of organisms (Malishev et al., 2018; Sears et al., 2016, 2019).

## Application of biophysical models to different types of organisms

*Ectotherms*

Most simply, biophysical models can be used to estimate body temperature of a single life stage of an ectotherm in a particular microclimate (A in Figure 5), such as lizard embryos laid at a given depth in the soil and under a specified level of shade (Levy et al., 2015). To identify constraints on species, body temperature predictions from biophysical models are typically combined with data on the temperature-dependence of development, sex, activity, growth, survival, or reproduction (Buckley, 2008; Mitchell et al., 2008). For example, Levy et al. (2015) combined predictions of lizard embryo temperatures through time at sites across the United States with laboratory data on lethal temperatures and the thermal dependence of development to determine whether the species could survive and develop at a particular site. Such analyses can be very useful for identifying areas and conditions where the species *cannot* persist. But it is often necessary to account for multiple life stages and for behavioral thermoregulation (B in Figure 5)—particularly when the organism's environment is highly heterogeneous —to gain a more complete picture of the constraining aspects of a species' fundamental niche. Additionally, the calculated potential activity time of the species at a site can be used to identify where activity restriction is likely to limit a species' distribution or abundance (C in Figure 5; (Buckley, 2008; Kearney, 2012; Levy et al., 2016a; Levy et al., 2017)). This can be done by assuming a fixed requirement for activity (Kearney & Porter, 2004), or by explicitly simulating energy and/or water intake using data on food properties and digestive physiology (D in Figure 5) and comparing these to modelled energy and water requirements (Buckley, 2008; Kearney et al., 2018).

*Endotherms*

Biophysical models for endotherms use the same principles, but usually assume a constant target body temperature (or a narrow tolerable range) and infer the consequences of this constraint for energy and water requirements (Porter et al., 1994). For a given body temperature, it is possible to solve for the metabolic rate that satisfies the energy balance equation (Figure 4). Under cold conditions, for example, the model can calculate the increase in energy expenditure needed to avoid hypothermia (Porter et al., 1994). Under warm conditions, the model can calculate the evaporative cooling costs needed to avoid hyperthermia, or the increase in body temperature in the absence of evaporation, assuming that the metabolic rate is constrained by a lower limit that represents the minimal rate of energy expenditure required for its current activity state (i.e., resting, digesting, moving) (Porter et al., 2000). Endotherms and ectotherms are generally modelled assuming they approximate a simple shape (e.g. sphere, ellipsoid) that has well-known heat-transfer properties (O'Connor & Spotila, 1992; Porter et al., 2000). However, multi-part models that are made up of various simple shapes (cylinders for legs, ellipsoid for the torso) have been used to better reflect the shapes of mammals and birds and to capture heat loss from appendages (Fitzpatrick et al., 2015; Mathewson & Porter, 2013). Moreover, animals with very complex geometries can have their convective heat exchange modelled using computational fluid dynamics (Dudley et al., 2013).

As with ectotherms, models of endotherms can be implemented in different ways, with different data requirements (Table 1). Most simply, they can be used to predict the energy or water costs of maintaining a set body temperature in a particular microclimate (McCafferty et al., 2011; Riddell et al., 2019; Southwick & Gates, 1975). Inferring distribution or activity limits for endotherms can be more difficult than for ectotherms because endotherm performance has a more complex response to temperature that is more dependent on water and food availability (but see Levy et al. (2016b; 2019); Mitchell (2018)). Data on food properties, intake and digestive physiology are needed for food and water balance calculations (Kearney et al., 2016; Porter et al., 2000). Therefore, most studies of endotherm distribution limits have focused on modelling a single life stage, usually adults. Several studies have incorporated the costs of lactation in mammals (Briscoe et al., 2016; Rogers et al., 2021) or estimated potential reproductive output (Kearney, et al., 2010c).

Table 1. Example studies that have used biophysical modelling to predict, attribute and understand species responses to climate change

| Global change question | Type of organism - Species | Life stages | Key model outputs (study area) | Key findings | Reference |
|---|---|---|---|---|---|
| **Attribute** observed shift in **phenology** to past climate change | Butterfly - *Heteronmypha merope* | Egg - adult (emergence) | Adult emergence date over time (single site - Melbourne, Australia) | Past climate change resulting in faster development of immature life stages explains the observed shift in timing of butterfly emergence across years | Kearney et al., 2010 |
| **Attribute** observed changes in **site occupancy** to past climate change | Birds - 50 species in the Mojave desert | Adults | Mapped cooling requirements i.e. required heat loss (Mojave desert, United States) | Increases in water requirements for evaporative cooling are positively associated with observed species declines, with further increases of 50-75% in cooling costs likely under future warming. Reductions in body size can reduce cooling costs, but are unlikely to offset projected increases under climate change. | Riddell et al., 2019 |
| **Predict** how climate change will alter **distribution** & **population dynamics** | Lizard - *Sceloporus undulatus* complex | Embryos and adults | Mapped rates of population growth (United States) | Impacts of climate change greatly underestimated if fail to account for vulnerable embryonic stage or use average monthly rather than daily climate data. | Levy et al., 2015 |
| **Predict** how climate change will alter **distribution** | Lizard - *Tiliqua rugosa* | Whole life cycle from egg (live bearing) to adult | Mapped outcomes of growth trajectories starting in different years, including activity, time to maturity, reproduction, life span, & ultimately the intrinsic rate of increase (Australia) | Water is a more potent limit on distribution than temperature and better explains current distribution. Future warming would benefit this species from a thermal point of view, but the incorporation of water constraints shows strong spatial variation in outcomes that depends on climate change scenario. | Kearney et al., 2018 |
| **Predict** how climate change will alter **activity times** | Rodent - *Acomys russatus* | Adults | Mapped energy and water costs of activity in different day parts (United States) | Warming will introduce or increase evaporative cooling costs in many locations, while decreasing shade and water availability. Diurnal animals can avoid evaporation cooling costs by switching to nocturnality. | Levy et al. 2019 |
| **Predict** how climate change will alter **migration routes** | Birds - *Alle alle* | Adults | Average energy requirements during migration and wintering phase (four alternative migration strategies from Franz Josef Land, Russian Federation) | The migration strategy with the highest flight costs (transarctic migration from the North Atlantic towards the North Pacific), is predicted to be half as costly, energetically, than the current migration strategy (migration to the North Atlantic) or high-arctic residency, because of more favourable conditions encountered on this route. | Clairbaux et al., 2019 |
| **Predict** current and future thermal **constraints** on **activity** | Grizzly bear - *Ursus arctos* | Adult - lactating & non-lactating females | Mapped metabolic rate predictions for lactating and non-lactating females with and without access to water (Yellowstone, United States) | The future distribution of grizzly bears in Yellowstone may be driven by individuals, particularly lactating females, access to water for cooling. | Rogers et al. 2021 |
| **Understand evolutionary responses** to past climate change | Butterfly - *Colias meadii* | Adults | Predicted fitness functions, directional selection, and evolutionary responses (a subalpine and alpine site in Colerado, United | Past climate warming has altered predicted patterns of directional selection, but climate variability limits predicted evolutionary responses. Approach captures trade-offs between trait values that optimise flight time and reduce risk of overheating, not anticipated | Kingsolver & Buckley 2015 |

| | | |
|---|---|---|
| | states, 1955-2010) | by simple theoretical models of responses to directional change. |

**What have biophysical models taught us so far?**

Biophysical models have long been applied to understand how climate constrains organisms and, more recently, to predict responses to future climate change (Figure 3a). Important lessons have emerged from these studies (see also Table 1). First, short-term weather conditions can strongly influence survival through time and space – processes that are not necessarily captured by annual or even monthly climate means commonly used in statistical models. For example, it is the combination of cold temperatures and high wind speed that results in high energy costs for wintering seabirds (Fort et al., 2009), while the combination of hot weather and low water availability/high humidity limits the distribution of the koala because individuals cannot lose sufficient heat via evaporative cooling and remain hydrated (Briscoe et al., 2016). Moreover, when using monthly means of soil temperatures, models may substantially underestimate lethal heat events that may kill lizards' embryos and lead to population declines (Levy et al., 2015). These studies show how important it is to get the temporal resolution right when inferring climate change impacts with biophysical models (Kearney et al., 2012).

We have also learned that the interaction between climate change and the seasonal availability of preferred thermal conditions is complex. For example, climate change is projected to lengthen the reproductive season of a North American lizard but with any fitness gains being offset by the negative impacts of warmer summers on embryo and juvenile survival (Levy et al., 2016a). Also, in high-elevation butterflies, dark-color adaptations that maximize absorbance of solar radiation may become maladaptive and reverse to decrease risks of overheating under climate change (Buckley & Kingsolver, 2019).

Studies have also repeatedly illustrated the importance of microclimates that protect individuals from high body temperatures or high rates of evaporative water loss. This includes deep shade (Kearney, et al., 2009a), cool underground refuges (Briscoe et al., 2022; Riddell et al., 2021), access to water for wallowing (Rogers et al., 2021), or cool leaves or tree trunks (Briscoe et al., 2014; Potter et al., 2009; Wolf et al., 1996). In the microclimatically complex intertidal zone, maximum mussel body temperatures were shown to have geographically varying sensitivity to a given increase in air temperature, with the body change always lower in magnitude than the air temperature change (Gilman et al., 2006). Energetic constraints have more severely impacted birds than small mammals in Death Valley over recent decades due to their lesser ability to shelter from climate change (Riddell et al., 2021) showing that species or populations that can exploit these microclimates may be less vulnerable to climate change (although see Buckley et al. (2015)). Protecting or providing these microclimates can thus be a useful target of management.

Biophysical modelling studies have also highlighted traits or behaviors likely to render species less vulnerable to climate change – such as flexible activity timing. Some animals can minimise their exposure to stressful conditions by altering their patterns of daily activity. Extensive sensitivity analyses focused on terrestrial ectotherms suggested that, of all the behavioral and physiological strategies available to them, a change to activity timing has the largest effect on predicted body temperatures (Stevenson, 1985b). Likewise, a study of diurnal rodent species predicted that a shift to nocturnal activity could compensate for the effects of climate change (Levy et al., 2019). However, there are often trade-offs between minimising thermoregulatory costs and avoiding lethal conditions and other activities, such as maximising food intake (Long et al., 2014).

Finally, studies have illustrated that different life stages have different environmental and/or nutritional requirements (Kingsolver et al., 2011); accounting for the whole life cycle is therefore important. For example, adult desert iguanas are predicted to be able to survive and grow at sites that do not have the right soil conditions (temperature and moisture) for their eggs to develop (Porter & Tracy, 1983). Additionally, there are often complex dependencies between life stages, meaning that the temporal pattern of conditions relative to the phenology of the animal is important (Briscoe et al., 2012).

## Limitations and opportunities

Models that explicitly capture mechanisms should, in principle, better predict organismal responses to global changes, but there remains a strong imbalance towards correlative approaches. Broadscale application of biophysically based mechanistic niche models to many species will require a large, concerted effort (Urban et al., 2022). Limitations in characterising environments, trait data collection and collation, education and software must be overcome but there are also exciting new opportunities to break through these limitations.

### Microclimates

Until recently, gathering input data for microclimate models involved searching, downloading and tailoring the relevant climate data. Recent implementations of R packages for microclimate modeling provide convenient access to online datasets such as NCEP (Kemp et al., 2012) and ERA5 (Klinges et al., 2022) climate datasets for use in microclimate models. The NicheMapR (Kearney & Porter, 2017) package allows users to specify a location and time window of interest, extract the input data, and run an expanded version of the Niche Mapper microclimate model (Porter et al., 1973). The microclima R package consists of functions for pre-adjustments of such input forcing data for important “mesoclimate” effects such as wind sheltering, coastal influences, cold air drainage, and elevation-associated lapse rates (Maclean et al., 2017, 2019). These two complementary packages have now been integrated (Kearney & Porter, 2020), highlighting the value of collaborations between research groups.

Another challenge is the relatively coarse temporal and spatial resolution of online climate databases compared to those of animals (Potter et al., 2013, Sears et al 2011). For many applications, hourly resolution data can be extrapolated from daily minimum and maximum values (Kearney et al., 2014). However, when organisms are sensitive to extreme temperatures or rare environmental combinations, hourly resolution is needed (Levy et al., 2015). Integrating high spatial resolution thermal landscapes with biophysical models can inform how organisms are constrained by thermal transients and trade-offs in their ability to access environments (Basson et al., 2017; Kearney, et al., 2021a; Malishev et al., 2018; Sears et al., 2016, 2019). Although such data are rare, recent developments in remote sensing can revolutionise microclimate estimates by capturing high- resolution data. For example, information from satellites, such as the LiDAR products of the Global Ecosystem Dynamics Investigation, and small drones can produce sub-meter resolution data (e.g., elevation, vegetation, thermal maps) (though see Maclean et al. (2021)).

Although microclimate models can supply quite accurate predictions for open habitats, more testing and development is needed to accurately model microclimates in habitats with complex vegetation and high spatial heterogeneity structure. For example, accounting for turbulence in forests remains challenging (Brunet, 2020), due to the complex interacting effects of vegetation, landscape, and wind on heat balance at fine resolutions. Heterogeneous environments are also challenging since it is hard to capture the strong non-linearities of the heat exchange components across the landscape. Although numerical methods can overcome these challenges (e.g. finite element approaches, see Baldocchi (1992) and Gastellu-Etchegorry et al. (2004)), they are computationally challenging. Fortunately, advances in data collection are substantially improving our capacity to validate models, or measure microclimate at complex landscapes where our models are still inaccurate (e.g., Fabbri & Costanzo, 2020).

Microclimate models involve many physical and numerical calculations. At higher spatial and temporal resolutions, these calculations may be too computationally intensive and require massive data storage facilities. These challenges often limit calculations to small geographic extents or to relatively coarse resolutions, and discourage storing and sharing of model outputs, increasing the need for repeated computation. There are at least three potential solutions to this problem. First, statistical models (Maclean et al., 2021) or Gaussian process emulation techniques (Conti et al., 2009) can estimate complex dynamics of microclimates over short time periods, eliminating the need to run microclimate models in time-increments and reducing run times. Second, modern computationally efficient programming languages, such as Julia (see below), offer the ease and expressiveness of high-level languages with performance comparable to Fortran or C++ (Bezanson et al., 2018). Finally, it may not always be necessary to model microclimate in a spatially explicit manner – often knowledge of the mean, variance and/or range of microclimatic conditions at a given locality may be sufficient to answer the research question (Bütikofer et al., 2020).

## Functional trait data

Biophysical models require detailed organismal trait data spanning morphology, physiology, behavior, and life history, to tailor predictions to specific taxa or questions. These are necessarily 'functional traits' because they act as important parameters or thresholds for models of an organism's performance (Dawson et al., 2021; Kearney et al., 2021b). Biophysical models are often criticised for being parameter-hungry (Buckley et al., 2010; Kearney & Porter, 2009), but with the rapidly increasing availability of trait databases, this criticism has become less valid.

Functional trait databases have developed rapidly for plants, with the number of entries for functional traits increasing from 2.07 to 11.85 million between 2007 and 2020 across nearly 280,000 plant species (Kattge et al., 2020), half of these being linked to specific geographic locations. Plant mechanistic models typically focus on growth rates as the primary metric of performance (Duursma & Medlyn, 2012; Schouten et al., 2020), though phenology is also commonly used (Chapman et al., 2014). However, there are still very few measurements of solar absorptivity (but see Gates (1980)).

Relative to plants, databases of functional traits for animals are less consolidated and extensive. Biophysical heat- and water-flux calculations of animals require on estimates of body size, area and shape as well as solar reflectance and emissivity. For endotherms, insulation properties (i.e., density, length, diameter of hairs) (Campbell & Norman, 1998; Gates, 1980) are required to estimate thermal conductivity of insulation, such as pelage or plumage, or conductivity can be measured directly from specimens (Porter et al., 1994; Riddell et al., 2021). Relevant physiological functional traits include basal or standard metabolic rate, cutaneous resistance to water loss, target body temperature, thermal tolerances and thermal optima. Behavioral traits include body temperature thresholds for thermoregulation, including thermoregulatory mode (or accuracy) and desiccation avoidance (Clusella-Trullas & Chown, 2014; Kearney, Shine, et al., 2009; Riddell et al., 2018; Sears et al., 2016). Gathering information on so many traits is challenging since functional trait databases for animals typically focus on one type of trait (Herberstein et al., 2022; Myhrvold et al., 2015), but more often, these traits are published for groups of animals (Bennett et al., 2018; Clusella-Trullas & Chown, 2014; Grimm, Annegret et al., 2014; Le Galliard et al., 2021; Madin et al., 2016; Oliveira et al., 2017). In addition to consolidated databases, there is a wealth of animal functional trait data available in the published literature, particularly for well-studied groups. For example, a literature search with terms relating to different types of relevant traits for lizards and snakes (see Supporting Information, Appendix 1) identified 9029 unique papers. Papers focused on thermal physiology were most common, followed by hydric physiology, morphology (excluding body mass), metabolism, and then studies that examined both behavioral thermoregulation and thermal physiology (Figure 3b).

Although the availability of functional trait data is rising, there is much room for improvement in how these data are collected and collated, and methodology can have a substantial impact on trait values. Thermal tolerances can exhibit important variation due to acclimation effects, the rate at which temperature changes, or the duration in which organisms are exposed to a temperature (Pintor et al., 2016; Sunday et al., 2019). Similarly, functional traits can vary depending on whether they are measured under constant or (more natural) fluctuating conditions (Morash et al., 2018; Niehaus et al., 2012). Species' traits can also exhibit substantial variation within and across populations, across developmental life stages, or in response to environmental cues over time (i.e., phenotypic plasticity) (Moran et al., 2016). Ensuring that individual-level measurements and relevant metadata are recorded in functional trait databases will help ensure that trait data are available and can reliably be used in biophysical modelling. For example, georeferenced trait data can be combined with environmental data to assess the environmental sensitivity of certain traits, while museum collections can be used to  quantify spatial and temporal variation in traits (Briscoe et al., 2015; Gardner et al., 2019). Alternatively, sensitivity analyses can assess the impact of variation in a particular trait, or identify the functional traits that most strongly influence estimates of performance (Augusiak et al., 2014; van der Vaart et al., 2016). A single general database of functional traits for biophysical (and metabolic) modelling would greatly enhance the uptake of the methods, make data deficiencies clear, and advance the study of functional traits in general (Kearney, Jusup, et al., 2021).

## Training and background

The lack of training in the requisite concepts and techniques is a substantial obstacle to widespread application of biophysical models in global change biology. Quantitative training in undergraduate ecological courses is often poor (Barraquand et al., 2014) and focused primarily on statistical approaches (Auker & Barthelmess, 2020). In our experience, modelling issues associated with dynamical systems models (Box 1), including the derivation of differential equations and their integration through time via numerical models, are alien to many modern biologists and ecologists. The availability of several open software packages for biophysical modelling means that implementing these methods is now easier than ever before. However, users still need to be familiar with the underlying principles and understand how these are implemented (including key simplifying assumptions and approximations) so that they can generate models appropriate to the question being asked (O'Connor & Spotila, 1992). This also requires a solid understanding of the natural history of the species being modelled.

Biophysical modelling draws on disparate fields, such as physics, engineering, climatology, physiology and behavioral ecology. Spending time becoming familiar with these topics and skills is a necessary and worthwhile investment for newcomers to this approach. A recommended starting point for those entering this field is to gain familiarity with the fundamental processes and equations describing basic forms of heat exchange as these form the bedrock of biophysical ecology. Interested readers are directed towards the free online educational resources created by a subset of the authors (*TrEnCh Project*, 2022), foundational textbooks (Campbell & Norman, 1998; Gates, 1980), and the online applications (*CAMEL*, 2022). Tutorials, vignettes and Shiny apps associated with R packages (NicheMapR, TrenchR) allow users to begin to practically apply these tools and become familiar with model parameters and outputs. Alongside further expansion of these online resources, greater exposure to biophysical modelling in undergraduate and postgraduate classes focused on physiology, ecology, and/or quantitative methods, as well as focused workshops and training opportunities at postgraduate-level and above, would help adoption.

Biophysical models require detailed morphological, physiological, behavioral and microclimatic data about the species under study parameterize and test them, and this data will often need to be collected by the researcher. While many measurements are relatively straightforward (e.g., measuring body mass and size, pelage depths or preferred temperatures), others can require specialist equipment or techniques (e.g., measuring solar reflectance, energy and water turnover using doubly-labelled water, or thermal dependence of metabolic rate or evaporative water loss using respirometry). Understanding and measuring the physical processes driving microclimates also requires discipline-specific expertise (Maclean et al., 2021). Collaborating with other researchers such as physiologists, meteorologists, hydrologists, or species experts, who have specialised equipment, expertise or existing data can help overcome these challenges. Indeed, greater collaboration between researchers in different fields would facilitate efforts to apply biophysical modelling more broadly, not only by enhancing data collection efforts but also by highlighting processes that may not be adequately captured by current models (Mitchell et al., 2018).

## Software ecosystems

The use of statistical programming tools – largely R, sometimes Python – has become ubiquitous for researchers of global change biology (Lai et al., 2019). This competence has developed in tandem with the emergence of vast *software ecosystems* that provide the many interoperable open-source packages we combine to process data and build models (Hoving et al., 2013; Plakidas et al., 2017). It is rare for researchers to write their own statistical algorithms: instead, statistical modellers combine freely available tools to analyse their specific problem using high-level model definitions. In statistics, modellers stand on the shoulders of thousands, across varied disciplines, who have published their tools on CRAN (Hornik, 2012) and contributed to R's software ecosystem.

Biophysical modelling requires researchers to write a different kind of software to the statistical scripts many are accustomed to. The differential equations of biophysical models often have heterogeneous, problem-specific structure, unlike the generic algorithms used in statistical approaches. This comes with a different set of social and technical problems to those encountered in statistical modelling and has limited the development of software ecosystems.

The situation is improving greatly in biophysical modelling, as tools like NicheMapR, microclima and TrenchR have been made open and available. But, in contrast to the broad base of contributors that statistical software draws on, we are only able to integrate existing code into a limited extent of our work. It is relatively common to use packages to provide microclimate or nutrient data to feed into custom metabolic models (see Supporting Information). However, it is rare to use existing packages as components to develop new, custom models. The outcome of this pattern is clear in the reviewed literature on animals: researchers of recent papers (> 2000) are divided in two groups, those parameterizing existing models, like the ectotherm model in NicheMapR (44%), and those writing custom models completely from scratch (49%). Some of the few cases of model modification directly edited the package code (~2%), a questionable practice for maintaining correctness and reproducibility. Biophysical modellers need more capacity to work between these extremes, the ability to easily modify only the required components of existing models, and generally to make better compromises between flexibility and effort.

There are several reasons for the current situation: most code was developed before open and reproducible coding principles; a preference for low-level languages like Fortran, that never developed significant software ecosystems; and under-recognised technical problems, e.g., that connections between components in biophysical models often needs to occur *inside* differential equations because processes often feedback on each other (e.g., leaf temperature and stomatal conductance in plants). Using package components inside differential equations often necessitates that the connections between model components (i.e., function calls) occur in high performance code, not in slower R or Python wrappers.

Biophysical modelling will likely continue to require high performance tools as the quantity and resolution of available data increases. But, to leverage previous work as statistical modellers do, biophysical modellers need modular tools that are also embedded in a software ecosystem and can be used together to construct new models without rewriting basic algorithms from scratch.

A potential solution to this problem is the Julia language (Perkel, 2019; Schouten et al., 2020, 2022). Julia has a rapidly growing, highly intercompatible software ecosystem targeted at scientific computing, differential equations and model optimisation. Its code is similar to dynamic languages like Python and R. However, it compiles packages and user scripts down to machine code at run-time giving performance comparable to Fortran. As an example of this potential, Julia is used for large scale biophysical modelling by the Climate Modelling Alliance (CliMA). The CliMA project combines model components maintained in separate repositories for their ocean, land and atmospheric models. Within the land model, specific tools for stomatal conductance and photosynthesis are defined in separate modular packages: notably these can be used independently from climate models for other kinds of biophysical research (Wang & Frankenberg, 2022).

An important outcome of relying on shared, generic tools better integrated into a software ecosystem, rather than custom scripts or tools from field-specific silos, will be that benchmarking and testing can be done across a larger number of researchers, to a higher standard. Interdisciplinary collaboration is also likely to improve from the process of using and developing shared tools.

Another important component of furthering mechanistic approaches in global change biology is developing computational infrastructure for model development and testing. The Ecological Forecasting Initiative has developed comprehensive infrastructure for near-term ecological forecasting that could readily be adapted for the mechanistic approaches that benefit longer-term forecasting (Dietze et al., 2018, 2021). Central to the computational infrastructure are databases with historical biological data for model testing and comparisons. The availability of relatively high-resolution historical climate and paleoclimate datasets means that it is now possible to revisit or reinterpret previous field studies or past extinctions, including using these data for model testing (Mathewson et al., 2017; Morris et al., 2022; Wang et al., 2018). An integral part of these future workflows will be generating and mapping realistic estimates of uncertainty, for example due to underlying climate forecasts, traits (including behavior) or model structure (Briscoe et al., 2016; Dietze, 2017).

## Vision for future of tackling global change biology problems

Our long-term vision for the future of biophysical modelling involves researchers, trained in the physical principles of biophysical ecology, using modular and flexible methods, and using data compiled in a standardized functional trait database, to answer diverse questions in global change biology. A barrier to realising this vision is that funding calls for projecting the biodiversity responses to climate change often seek applied projections for many species, analogous to that feasible with statistical models (i.e., correlative species distribution models). In contrast, furthering models built around biophysical ecology will require extensive basic research and investment in the initiatives outlined above. We argue such efforts are nonetheless essential to adequately projecting biological responses to environmental change.

As biophysical models become more widely used, the accumulation of case studies from different systems, as well as improved infrastructure for testing and comparison, will aid in finding a middle ground whereby predictions include sufficient biological mechanisms for accuracy but are feasible to implement and facilitate further uptake of the methods. Additionally, detailed models implemented and tested for varied taxa in limited locales can achieve some generality by identifying important limiting mechanisms that can be investigated for other organisms in other locations. For example, many existing biophysical models of ectotherms focus on responses to temperature because the mechanistic basis of temperature responses are best empirically probed and understood and trait data are more available (Figure 3b). However, studies that explicitly incorporate water balance and how this constrains behavior have highlighted the importance of these processes, and provide templates for incorporating these aspects (Kearney et al., 2013, 2018; Riddell et al., 2017). Given the prominence of multiple stressors as climates change (Gunderson et al., 2016), it will be important to more routinely account for interactions between stressors such as water and oxygen balance, and to consider the dynamics of whole life-cycles (Kingsolver et al., 2011; Porter & Tracy, 1983). Investigation of hypotheses such as the oxygen- and capacity-limited thermal tolerance can also inform the expansion of biophysical models (Pörtner, 2021).

As we have discussed, biophysical models can be used alone or can be incorporated into other models that capture key processes – including movement, population dynamics, biotic interactions, and evolution (Buckley et al., 2010; Urban et al., 2016, 2022). Indeed, many of the earliest biophysical modelling studies directly incorporated these latter processes (Dunham & Overall, 1994; Kingsolver, 1979; Porter et al., 1973). With modular, general, biophysical modelling software, such studies will become more feasible, supporting the development of integrated mechanistic biodiversity models (Urban et al., 2022). For example, estimates of survival and/or potential reproduction from biophysical models can be used as inputs in spatially explicit population dynamics models, to better capture how biophysical processes combine with demographic traits to constrain population growth (Buckley et al., 2008). Likewise, predictions of energy and water costs associated with different environments can be integrated into individual-based models that explicitly model behavior as the outcome of trade-offs between factors such as thermal and hydric costs, food and water intake, predation risk, competition, and social activities (Malishev et al., 2018; Sears et al., 2016). Such approaches may be particularly important for accounting for missed opportunity costs in climate change forecasts (Cunningham et al., 2021). Species responses to environmental change are likely to be strongly driven by biotic interactions (Buckley, 2013; Jankowski et al., 2010). For example, incorporating likely changes in bamboo distribution exacerbates the predicted effect of climate change on the giant panda (*Ailuropoda melanoleuc*a) (Zhang et al., 2018).

An often discussed but seldom implemented approach to expanding biophysical modelling approaches is “hybrid” models, which use computational pattern-based approaches to inform uncertain or unknown parameters or relationships (Buckley et al., 2010; Dormann et al., 2012). The most common strategy is to include mechanistically derived layers (such as potential activity durations, heat-units available for development, incidence of stressful environmental conditions, or energy balances) as predictors in correlative species distribution models (Mathewson et al., 2017; Mi et al., 2022). While these methods are still closer to the statistical end of the spectrum (i.e., the data lead the dance), using mechanistically derived layers that translate timeseries of environmental conditions into metrics of fitness relevant to the species should help these models predict more reliably to novel conditions. Alternatively, Bayesian statistics and "domain-aware" or "model-informed" machine learning models can be used to inform statistical models with biological information and constraints, which can come from biophysical models or experimental results (Beery et al., 2021; Kotta et al., 2019). Additionally, inverse modelling could be used to infer biophysical model parameters from endpoints such as occurrences (Evans et al., 2016; Fordham et al., 2022).

One important motivation for furthering biophysical models is that they can readily address global changes such as the spread of invasive species and diseases and habitat loss or degradation – and how these will interact with future climate change. For example, the ability to generate forecasts in novel environments and identify management levers means that biophysical models are particularly useful for modelling invasive species (Barton & Terblanche, 2014; Chen et al., 2021; Kearney et al., 2008). Similarly, they have been used to map the spread of diseases such as chytrid fungus in the Northern cricket frog (*Acris crepitans*), where relationships between infection prevalence and/or survival with body temperature are known (Sonn et al., 2020). Biophysical models can also be integrated into scenario modelling to assess how different forms of global change (e.g., land use, climate change) will alter species distribution or population dynamics in the future (Nowakowski et al., 2017).

Overall, as biophysical models become more integrated into studies of global change, we will develop stronger linkages between physical and biological disciplines, greater predictive capacity, and greater understanding of the relevant processes and how to manage them.